\documentclass[12pt]{article}
\usepackage{epsfig}
\input epsf.sty
\newcommand{\la}[1]{\label{#1}}
 
\newcommand{\be}{\begin{equation}}
\newcommand{\ee}{\end{equation}}
\newcommand{\ba}{\begin{eqnarray}}
\newcommand{\ea}{\end{eqnarray}}
\newcommand{\bi}{\begin{itemize}}
\newcommand{\ei}{\end{itemize}}

\newcommand{\nr}[1]{(\ref{#1})}
\newcommand{\tr}{{\rm Tr\,}}

\newcommand{\fr}[2]{{\frac{#1}{#2}}}
\newcommand{\msbar}{\overline{\mbox{\rm MS}}}
\newcommand{\lambdamsbar}{\Lambda_{\overline{\rm MS}}}

\newcommand{\RR}{{\rm I\kern -.2em  R}} 
\newcommand{\eq}{Eq.~}

\newcommand{\fig}{Fig.~}
\newcommand{\figs}{Figs.~}
\newcommand{\se}{Sec.~}
\newcommand{\half}{{1\over2}}

\def\lsi{\raise0.3ex\hbox{$<$\kern-0.75em\raise-1.1ex\hbox{$\sim$}}}
\def\gsi{\raise0.3ex\hbox{$>$\kern-0.75em\raise-1.1ex\hbox{$\sim$}}}
\newcommand{\lsim}{\mathop{\lsi}}
\newcommand{\gsim}{\mathop{\gsi}}

\makeatletter \@addtoreset{equation}{section} \makeatother

\makeatletter
\renewcommand\section{\@startsection {section}{1}{\z@}%
                                   {-5.5ex \@plus -1ex \@minus -.2ex}
                                   {2.3ex \@plus.2ex}%
                                   {\normalfont\large\bfseries}}
\renewcommand\subsection{\@startsection{subsection}{2}{\z@}%
                                     {-3.25ex\@plus -1ex \@minus -.2ex}%
                                     {1.5ex \@plus .2ex}%
                                     {\normalfont\normalsize\bfseries}}
\renewcommand\thesection {\@arabic\c@section}
\renewcommand\thesubsection   {\thesection.\@arabic\c@subsection}
\renewcommand{\@seccntformat}[1]{%
\csname the#1\endcsname.\hspace{1.0em}}
\makeatother

\begin{document}
 
\begin{titlepage}
\begin{flushright}
CERN-TH/2002-169\\ 
NSF-ITP-02-43 
\end{flushright}

\begin{centering}

\vspace{0.6cm}
 
{\bf FINITE TEMPERATURE Z($N$) PHASE TRANSITION\\ 
WITH KALUZA-KLEIN GAUGE FIELDS}

\vspace{0.6cm}
 
K. Farakos$^{\rm a,}$\footnote{konstadinos.farakos@cern.ch}, 
P. de Forcrand$^{\rm b,c,}$\footnote{forcrand@phys.ethz.ch}, 
C.P. Korthals Altes$^{\rm d,c,}$\footnote{chris.korthal-altes@cpt.univ-mrs.fr},
\\
M. Laine$^{\rm c,e,}$\footnote{mikko.laine@cern.ch}, and
M. Vettorazzo$^{\rm b,}$\footnote{vettoraz@phys.ethz.ch}

\vspace{0.3cm}

{\em $^{\rm a}$%
Physics Department, NTU, 
15780 Zografou Campus, Athens, Greece\\}

\vspace{0.3cm}

{\em $^{\rm b}$%
Institut f\"ur Theoretische Physik, ETH Z\"urich,
CH-8093 Z\"urich, Switzerland\\}

\vspace{0.3cm}

{\em $^{\rm c}$%
Theory Division, CERN, CH-1211 Geneva 23, Switzerland\\}

\vspace{0.3cm}

{\em $^{\rm d}$%
Centre Physique Th\'eorique, CNRS, Case 907, Luminy,
F-13288 Marseille, France\\}

\vspace{0.3cm}

{\em $^{\rm e}$%
ITP, University of California, Santa Barbara, CA 93106-4030, USA}

\vspace*{0.6cm}
 
\end{centering}
 
\noindent
If SU($N$) gauge fields live in a world with a circular extra
dimension, coupling there only to adjointly charged matter, the system
possesses a global Z($N$) symmetry. If the radius is small enough
such that dimensional reduction takes place, this symmetry is 
sponta\-neously broken.  It turns out that its fate
at high temperatures is not easily decided with
straightforward perturbation theory. Utilising non-perturbative
lattice simu\-lations, we demonstrate here that the symmetry does get
restored at a certain temperature $T_c$, 
both for a 3+1 and a 4+1 dimensional world (the latter with
a finite cutoff). To avoid a cosmological domain wall problem, 
such models would thus be allowed only if the
reheating temperature after inflation is below $T_c$.
We also comment on the robustness of this phenomenon with
respect to small modifications of the model.
%
 

\vspace*{0.6cm}
 
\noindent
CERN-TH/2002-169\\ 
NSF-ITP-02-43\\ 
December 2002

\vfill

\end{titlepage}

\setcounter{footnote}{0}

\section{Introduction}

It has recently been suggested that new venues for, e.g., Grand 
Unification, could be obtained by considering compact extra dimensions 
of an inverse radius at the TeV scale, with non-Abelian gauge fields
propagating in the bulk~(see, e.g.,~\cite{ia,ddg,moose} and references
therein).

It is well known that if SU($N$) gauge fields 
were to couple only to adjointly 
charged matter, and the radius of a compact circular dimension
is such that a weak coupling
computation is reliable, then the system possesses a spontaneously
broken global Z($N$) symmetry, related to the Polyakov loop in the
extra direction~\cite{gh,gpy,nw,hosotani}.  This leads to the
existence of topological domain wall defects~\cite{lhk,nhs,bk,ms}.

Of course, such domain walls would
carry a huge energy density, $\sim $TeV$^3$,
and could thus not be produced under any normal circumstances.
However, if
the symmetry gets restored at a high temperature $T_c$, domain walls could
possibly be produced in a cosmological phase transition in the Early
Universe.  On the other hand, the existence of domain walls with
energy densities exceeding about MeV$^3$ can be excluded
experimentally, for instance, through the anisotropies in the cosmic
microwave background radiation~\cite{z,vs}.  This means that either
particle physics models of this type are not realised in nature or, in
analogy with how the monopole problem can be avoided in the case of
traditional grand unified theories, 
the cosmological history is such that the
reheating temperature after inflation is
sufficiently below $T_c$.

Given the meaningful conclusions that can be drawn, it seems
worthwhile to ask whether there really is a symmetry restoring phase
transition in these models. This is not automatically guaranteed to be
the case~\cite{sw}: there are even non-perturbatively established
examples of broken symmetries which do not get restored at high
temperatures~\cite{inv}. Then one would obviously have no domain wall
problem~\cite{ds}\footnote{Another possible loophole might be if the
    field responsible for the domain wall is very weakly coupled to the 
    plasma~\cite{cs}. In our case the relevant field is $A_y$, the
    component of the gauge field in the extra direction, and it couples
    strongly enough even to stay in thermal equilibrium.}.  

It turns out that in the present context the issue cannot easily be solved
with perturbation theory, since it breaks down close to the phase
transition point~\cite{kal}.  On the other hand, various
non-perturbative arguments suggest that the symmetry should get
restored~\cite{kal}.  Here we confirm the latter arguments with
lattice simulations.

The model we study is pure SU(2) gauge theory. We take it to live
either in four Euclidean dimensions (4d), standing for one time
direction, two space directions, and one compact extra dimension ---
or in five dimensions (5d), standing for one time direction, three
space directions, and one compact extra dimension. A continuum limit can
only be approached in 4d, since a 5d pure
Yang-Mills theory is not renormalisable.
Nevertheless, for a fixed lattice spacing,
the qualitative features of the phase diagram can be addressed also in
the 5d case.

In~\se\ref{se:lattice} we specify the lattice model studied, as well
as the observables measured.  We present our 4d data
in~\se\ref{se:four}, our 5d data in~\se\ref{se:five}, and conclusions
in~\se\ref{se:concl}.

\section{Basic formulation}
\la{se:lattice}

\subsection{Lattice action and observables}

We denote the number of dimensions by $d$, $d=4,5$.  Two of the
dimensions are assumed compact and periodic: a time direction $\tau =
0...L_\tau$, where $T = L_\tau^{-1}$ is the temperature, and a compact
spatial dimension $y=0...L_y$, where $M \equiv L_y^{-1}$.
The lengths of the remaining (in principle infinite) dimensions are
denoted by $L_i$, $i=1,...,d-2$.

We regularise the theory by introducing a spacetime lattice, with a
lattice spacing $a$. The lattice spacing is taken the same in all
directions. We denote $L_\tau = a N_\tau$, $L_y = a N_y$, $L_i = a
N_i$.  The lattice action is of the standard Wilson form,
\ba
 S & = & \beta_G \sum_x \sum_{\mu<\nu} \Bigl( 1-\fr12 \tr
 P_{\mu\nu}\Bigr),
\ea
where $P_{\mu\nu}$ is an SU(2) plaquette, and $\mu,\nu = 1,...,d$. The
(dimensionless) value of $\beta_G$ determines the lattice spacing $a$;
we return to the conversion in~\se\ref{phys_4d}.

Our aim, from the point of view of the discretised theory, is to
determine the phase diagram of the system in the three-dimensional
space spanned by $\beta_G,N_\tau,N_y$. The spatial volume $V \equiv
L_1 L_2 ... L_{d-2}$ is assumed extrapolated to infinity.

The observables we employ in order to determine the phase diagram are
the Polyakov loops in the $\tau$ and $y$ directions,
\ba
 P_\tau(x_i,y) & = & \fr12 \tr \prod_{n=0}^{N_\tau-1} U_\tau
 (x+n\hat{\tau}), \\ 
 P_y(x_i,\tau) & = & \fr12 \tr \prod_{n=0}^{N_y-1} U_y
 (x+n\hat{y}),
\ea
where the $U_\tau,U_y$ are link matrices, $x=(x_i,\tau,y)$, 
$i=1,...,d-2$, and
$\hat{\tau}$, $\hat{y}$ are unit vectors in the directions of
$\tau,y$.  We also monitor the plaquette, 
$\langle \tr P_{\mu\nu}(x) \rangle$.

For both observables we measure the distribution of their volume
average, 
\be
 \bar P_\tau \equiv \frac{a^{d-1}}{V L_y } \sum_{x_i,y}
 P_\tau(x_i,y), \quad
 \bar P_y \equiv \frac{a^{d-1}}{V L_\tau } \sum_{x_i,\tau}
 P_y(x_i,\tau) . 
\ee
Let us define, in particular,
\be
 |P_\tau| \equiv  \langle \,| \bar P_\tau | \,\rangle \,,  \quad
 \chi(P_\tau) \equiv \frac{V L_y }{a^{d-1}} \Bigl[ 
 \langle \, (\bar P_\tau)^2 \,\rangle - 
 \langle \, | \bar P_\tau  |\,\rangle^2 \Bigr] \,,  \la{OPy} \la{OXy}
\ee 
and in complete analogy for $|P_y|, \chi(P_y)$.
Physically, the role of $|P_y|$ is to tell whether
dimensional reduction into a $d-1$ dimensional effective theory takes
place ($|P_y| > 0$) or not ($|P_y| = 0$), while the role of $|P_\tau|$ is
to tell whether the effective $d-1$ dimensional theory is confining
($|P_\tau| = 0$) or not ($|P_\tau| > 0$). Let us recall that
here $d-1$ is supposed to
comprise the $d-2$ flat physical spatial dimensions, as well as the
one time dimension; for $d=4$ we are thus 
in effect considering a toy model of a
(2+1)-dimensional world.

\subsection{Universality class}
\la{se:univ}

As we vary $\beta_G,N_\tau,N_y$, we may expect to find phase
transitions.  In case the transitions turn out to be of the second
order, as is mostly the case, 
they are associated with some universality class, which we now discuss.

We have introduced two actual order parameters for the system,
${P_y}$, ${P_\tau}$, whose phases are Z(2) variables. 
In the limit of $N_\tau\gg N_y$ or $N_y\gg N_\tau$, only
one of them is ``critical'' at the transition point; then the
scaling goes to a good approximation according to 
the Ising model in $d-1$ dimensions~\cite{sy}.  
For comparable $N_\tau, N_y$ when both can be 
critical, the universality class is that
of the Z(2)$\times$Z(2) model in $d-2$ dimensions.  The properties of this
model however depend on a number of parameters, defining the relative
strengths of the self-interactions of the two spins, as well as the
interactions between them. In general the transition
is still of the second order, but in special cases it can also be of
the first order, at least for $d-2=3$~\cite{ruda}. Below, we will
indeed encounter first order phase transitions in the 5d case. 

\subsection{Perturbative predictions}
\la{se:pert}

Before turning to lattice results, let us also briefly review the
predictions of continuum perturbation theory.  The 1-loop finite
temperature continuum effective potential for the phase of the
Polyakov line, $P_y$, has been computed in~\cite{kal}, 
both for $d=4,5$. The structure
of the potential is such that at zero temperature, assuming weak
coupling for the given value of $M$ (say, $M \gg \lambdamsbar$ in the
4d case), the Z(2) symmetry related to $P_y$ is broken~\cite{bk}, 
such that $|P_y| > 0$. This is consistent with the fact 
that the theory undergoes dimensional reduction to $d-1$
dimensions~\cite{adjoint}.

Increasing now the temperature, the potential turns out to show no sign of
symmetry restoration. However, at temperatures parametrically of the
order $T \sim M/g^2$, perturbation theory is seen to break
down~\cite{kal}, and non-perturbative arguments can be given that the
symmetry should get restored~\cite{kal}.  The resulting phase diagram
was sketched in~\fig2 of~\cite{kal}.

\section{Lattice results in four dimensions}
\la{se:four}

In order to study the phase diagram with lattice simulations, we shall
employ various fixed values of $N_y$, $N_y = 1,2,4$. In each case, we
study a number of different $N_\tau$,
and determine the phase diagram as a function of
$\beta_G$.  We plot the results in the plane ($1/N_\tau,\beta_G$).  For
a fixed $\beta_G$ (fixing $a$), increasing $1/N_\tau$
(moving right) corresponds then to increasing the temperature, since
$1/N_\tau = a T$.

\begin{table}[t]

\centering
\begin{tabular}{lll}
\hline
 $N_\tau$ & $N_y$ & $N_1\times N_2$ \\ \hline 
 1,2,4 & 1 & $16^2$ \\
 1...6 & 2 & $16^2$ \\ 
 1...6 & 4 & $12^2,16^2$ \\ \hline
\end{tabular}

\caption[a]{The volumes studied in the 4d case. Note that on account
of the symmetry $N_\tau\leftrightarrow N_y, P_\tau \leftrightarrow
P_y$, some of them are effectively listed twice. }

\la{table_4d}
\end{table}

The 4d simulation points are shown in~Table~\ref{table_4d}.  
At every $\beta_G$ close to a phase transition, 
we study five $\beta_G$-values more carefully, each with a statistics of 40K
measurements (separated by 1 Metropolis and 6 overrelaxation
sweeps). 10K sweeps are discarded for thermalization.  Interpolation
in $\beta_G$ is achieved through Ferrenberg-Swendsen
reweighting~\cite{fs}, with jackknife error estimates.  The location
of the susceptibility maximum is obtained with a parabolic interpolation
through the binned, reweighted data.

\subsection{Existence of phase transitions} 

\begin{figure}[tb]

\centerline{~~\begin{minipage}[c]{6.4cm}
    \psfig{file=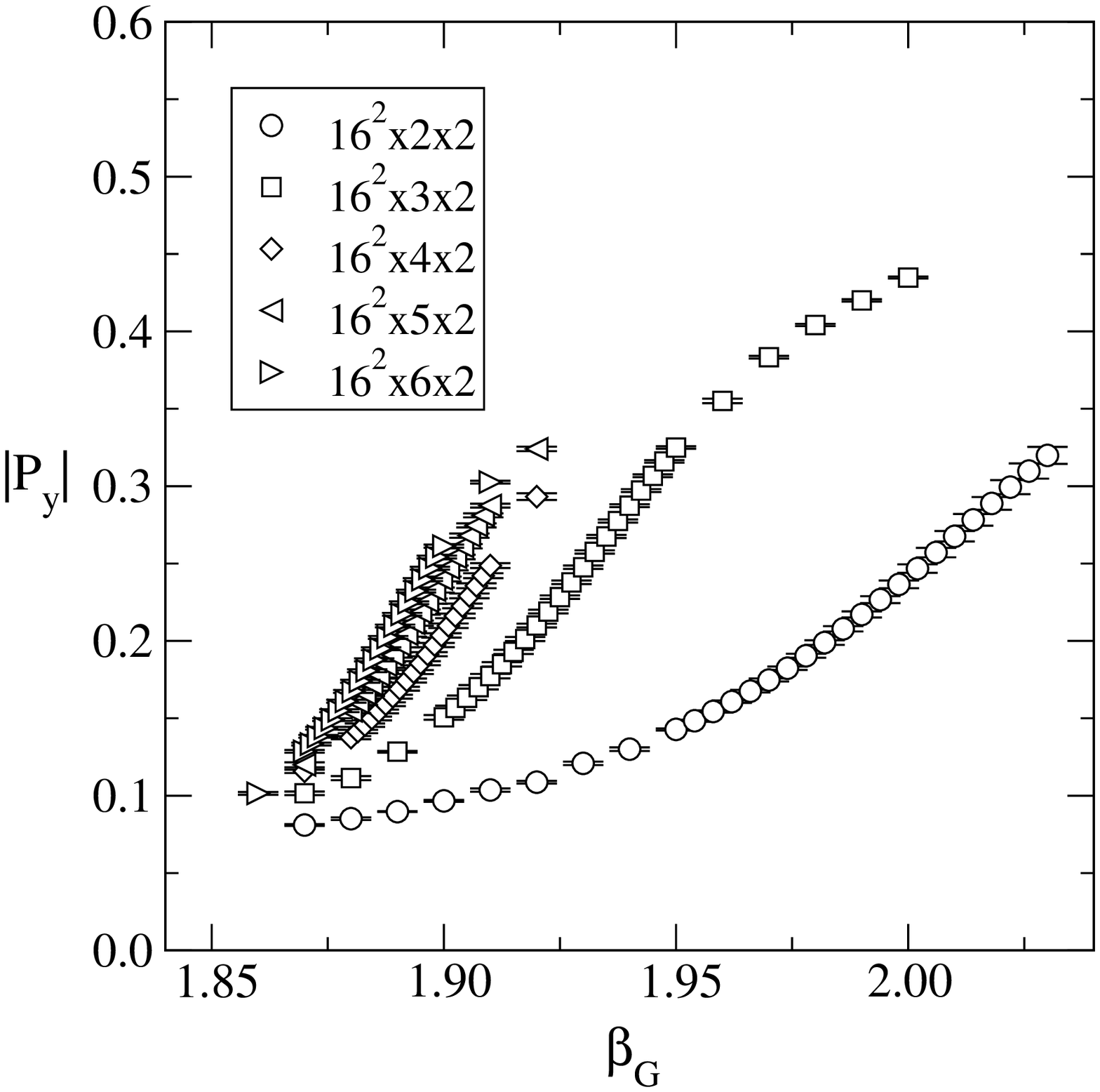,angle=0,width=6.4cm} \end{minipage}%
    ~~~~~\begin{minipage}[c]{6.4cm}
    \psfig{file=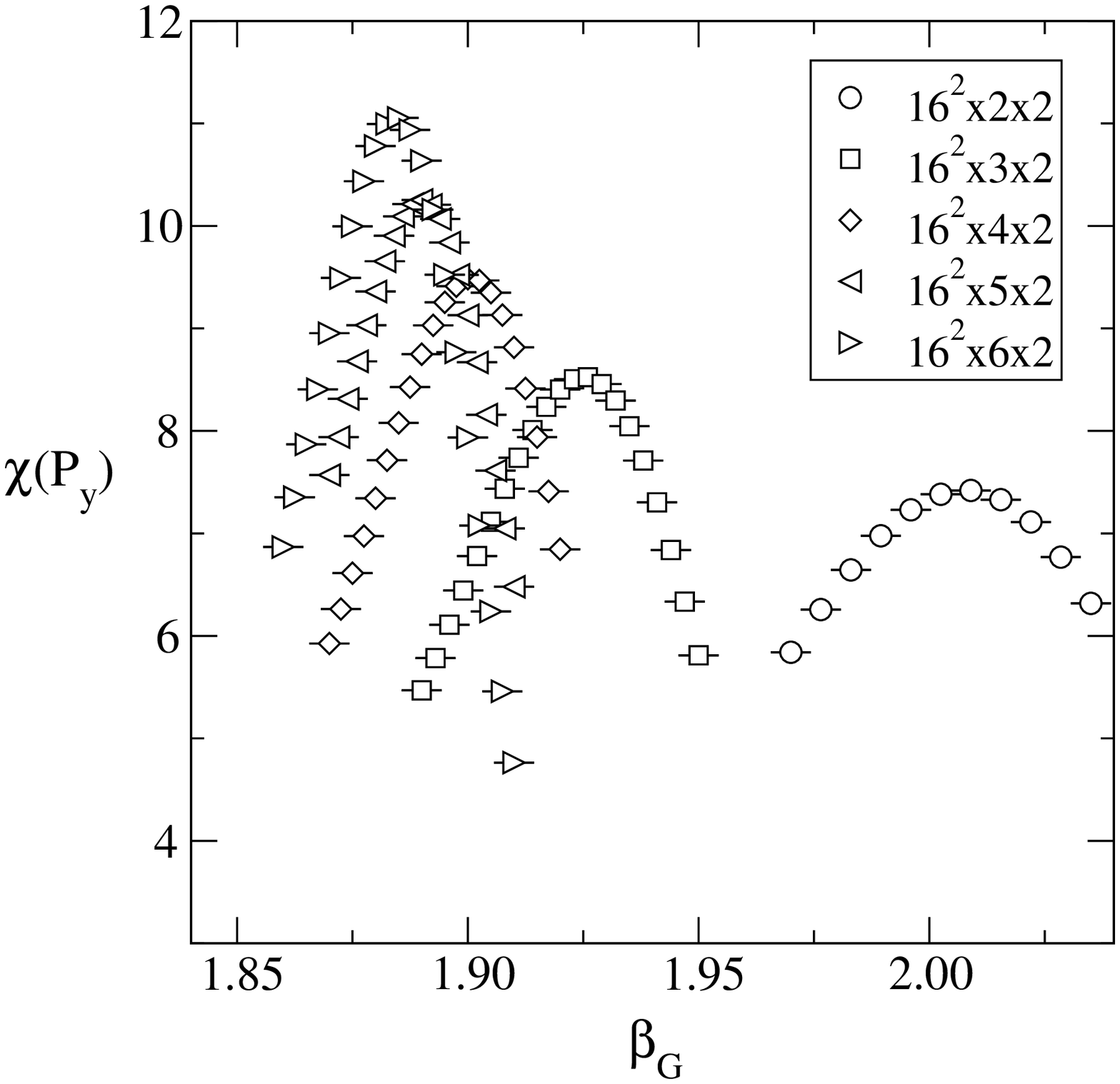,angle=0,width=6.4cm} \end{minipage}}

\vspace*{0.5cm}

\caption[a]{Left: the observable $|P_y|$ defined in~\eq\nr{OPy}, 
for $N_y=2$, and various $N_\tau$.  Right:
the corresponding susceptibility $\chi(P_y)$, defined
in~\eq\nr{OXy}. For $V\to\infty$, $|P_y|\to 0$ for $\beta_G$ to the 
left of the susceptibility maximum 
(cf.~\figs\ref{5d_second}, \ref{5d_first}).}

\la{fig:OPy_4d}
\end{figure}

In \fig\ref{fig:OPy_4d} we show the results obtained for ${|P_y|}$, as
well as the susceptibility $\chi({P_y})$, as a function of $\beta_G$,
at various $N_\tau$, for a fixed $N_y = 2$.  We observe that $|P_y|$
grows rapidly (symmetry breaks) when $\beta_G$ is increased beyond a
certain point. The transition point is located from the maximum of
$\chi(P_y)$. We see that the location of the phase transition depends
on $N_\tau$. Measurements of ${|P_\tau|}$, $\chi{(P_\tau)}$ show
similar phase transitions, however in general at some other
$\beta_G$. By studying two different volumes for $N_y = 4$
(Table~\ref{table_4d}), we know that the results are already
representing the thermodynamic limit  well enough for our purposes, 
as far as the position of the
transition is concerned (see~\fig\ref{fig:unphys_4d}, bottom
right). For $d=5$ (\se\ref{se:five}), we have carried out somewhat more
extensive volume scaling tests, and they do not give rise to any
concerns about our resolution.

\subsection{Phase diagrams for fixed $N_y$}
\la{unphys_4d}

\begin{figure}[tb]

\centerline{~~\begin{minipage}[c]{6.4cm}
    \psfig{file=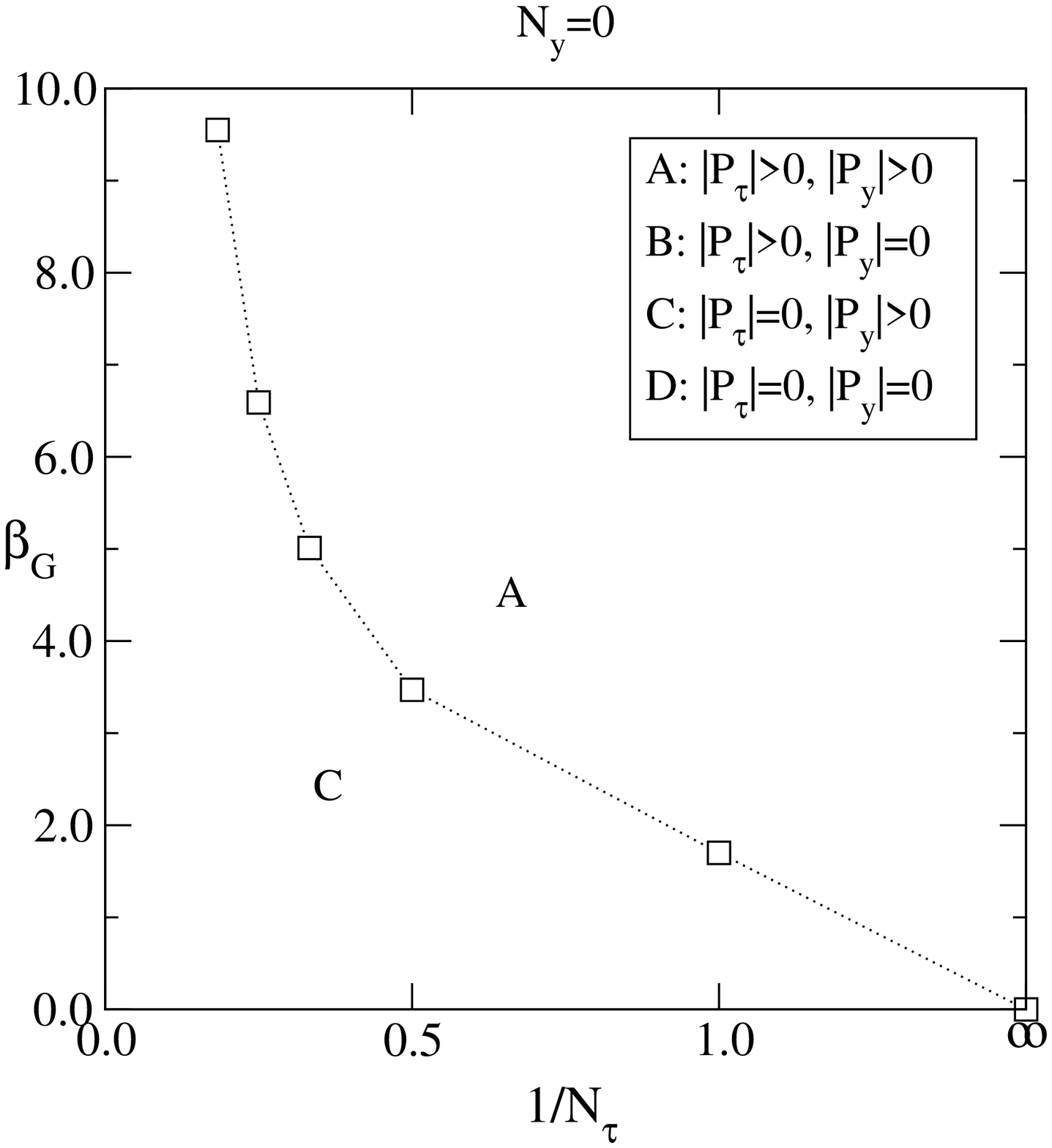,angle=0,width=6.4cm} \end{minipage}%
    ~~~\begin{minipage}[c]{6.4cm}
    \psfig{file=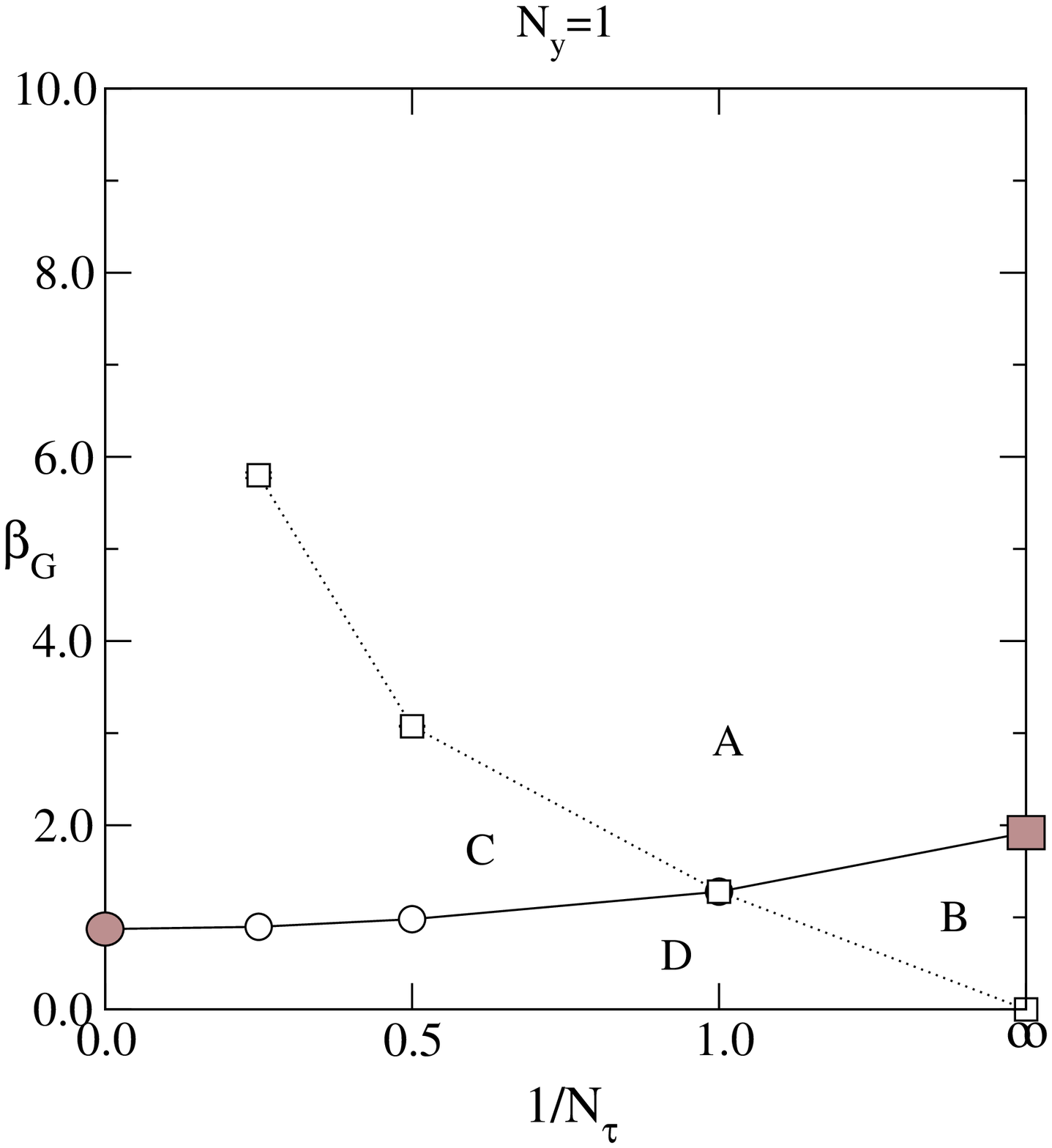,angle=0,width=6.4cm} \end{minipage}}

\vspace*{0.5cm}

\centerline{~~\begin{minipage}[c]{6.4cm}
    \psfig{file=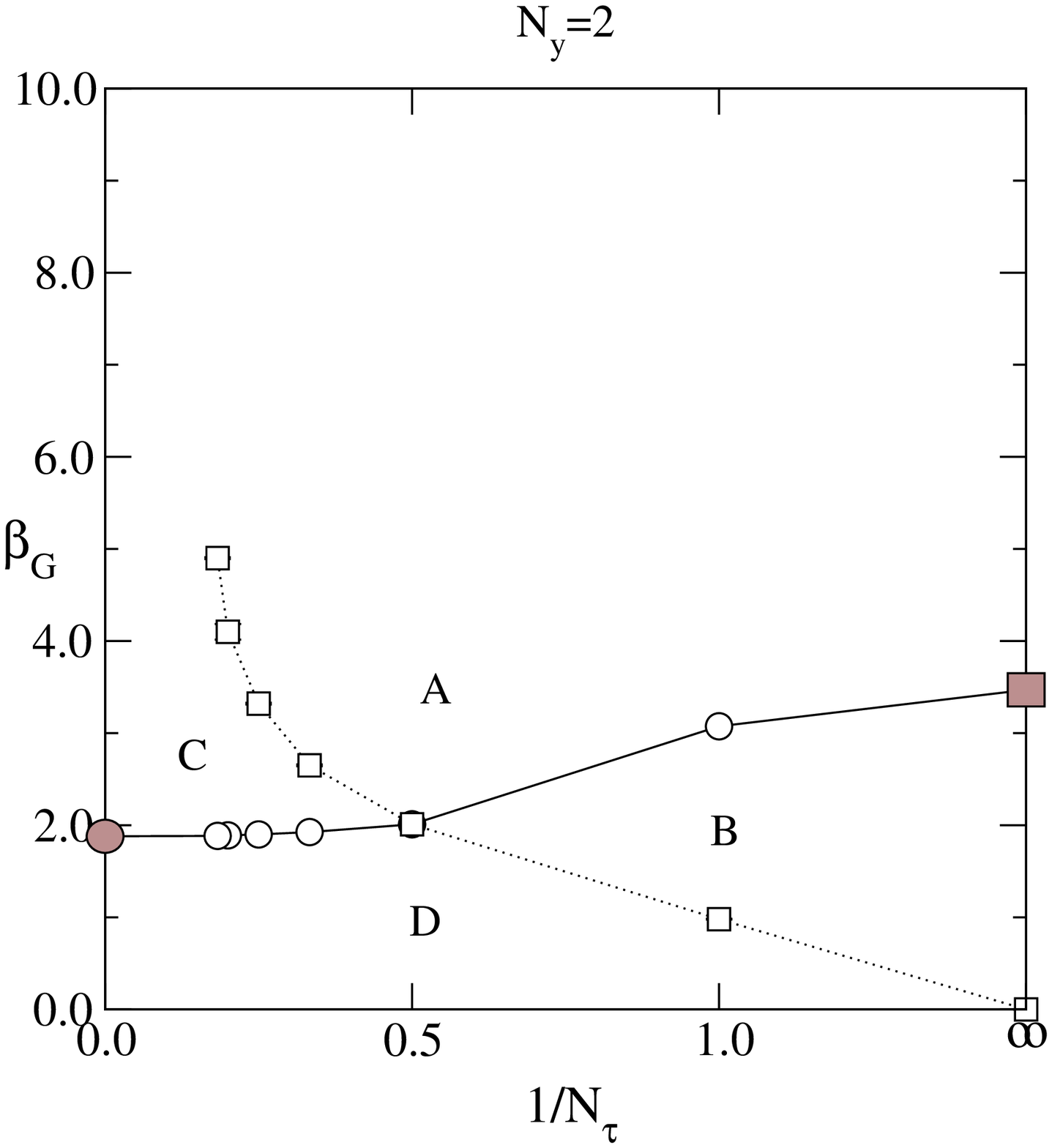,angle=0,width=6.4cm} \end{minipage}%
    ~~~\begin{minipage}[c]{6.4cm}
    \psfig{file=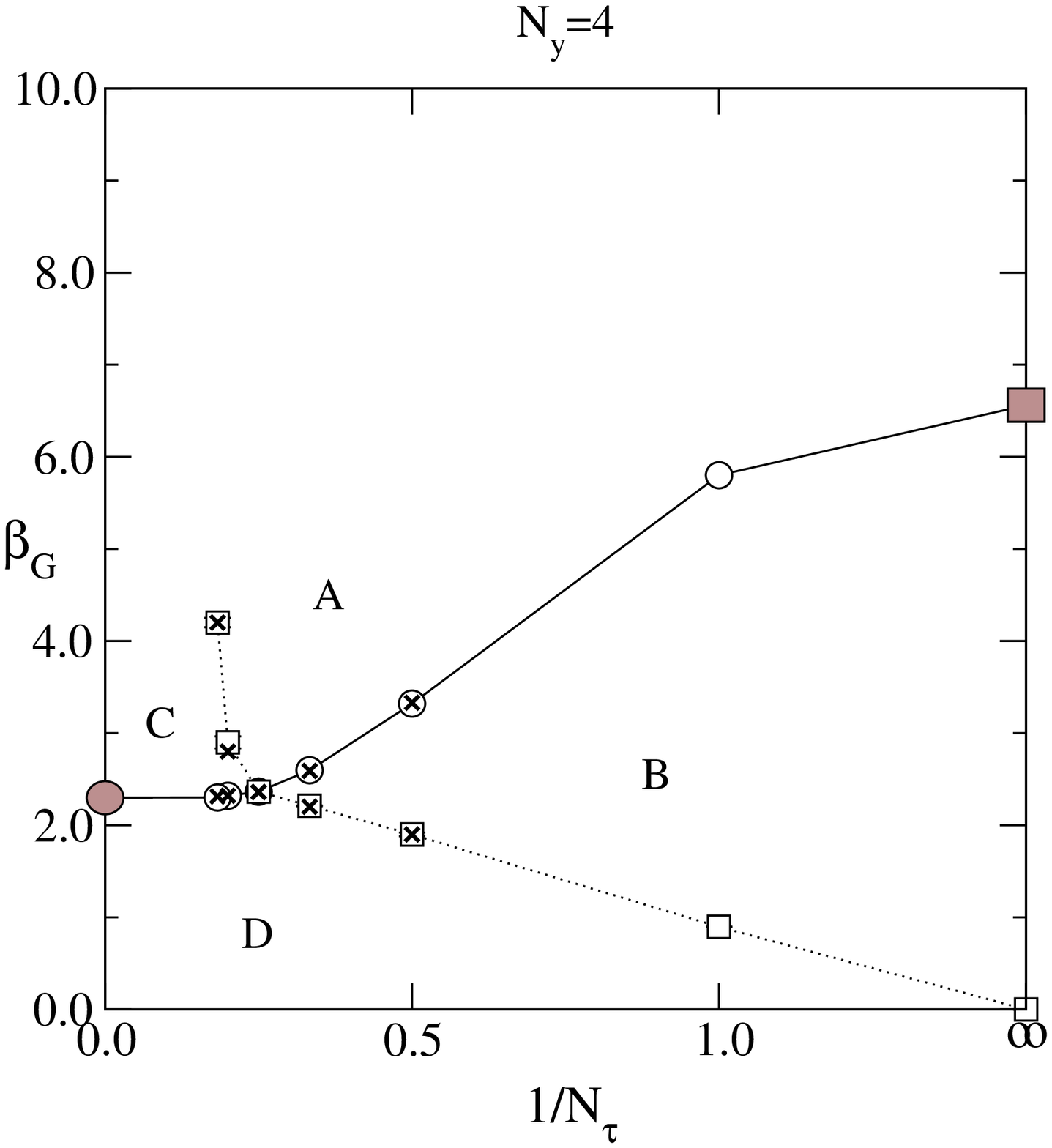,angle=0,width=6.4cm} \end{minipage}}

\vspace*{0.5cm}

\caption[a]{Phase diagrams for 
various $N_y$, in 4d. The lines connecting the simulation points are
there to guide the eye only.  The data for $N_y=0$ at $N_\tau \ge 2$, 
as well as that at $1/N_\tau = \infty$ (filled boxes),
are from the (2+1)d studies in~\cite{mt,eklllps}, while
the limiting values at $1/N_\tau = 0$ (filled circles) are
from the (3+1)d studies in~\cite{bems,fhk}. For $N_y=4$ we have shown
results both at the spatial volume $12^2$ (small crosses) and $16^2$
(large open symbols), finding no appreciable volume dependence.}

\la{fig:unphys_4d}
\end{figure}

When results such as discussed above are
collected together, we obtain the 4d phase diagrams shown
in~\fig\ref{fig:unphys_4d}.

As a first check, we may compare our results with some known
limits in the literature. The critical values of $\beta_G$ for a
(2+1)d SU(2) theory at finite temperatures have been determined
in~\cite{mt,eklllps}.  These results are collected
in~\fig\ref{fig:unphys_4d}, top left, showing the transitions felt by
${|P_\tau|}$ in the limit $N_y=0$, but they also determine the
transition points felt by ${|P_y|}$ at $1/N_\tau =
\infty$ (filled boxes).  The other known
limit is obtained from (3+1)d at finite temperatures~\cite{bems,fhk}:
these results fix the transitions felt by ${|P_y|}$ at $1/N_\tau =0$
(filled circles). Our data indeed interpolate between
these known values.

A second check may be obtained by considering the behaviour of the
transition for ${|P_y|}$ in the vicinity of $1/N_\tau =0$. Indeed, the
behaviour there is sensitive to the universality class of the
system. The point $1/N_\tau =0$ represents a (3+1)d theory, so that
the transition should belong to the 3d Ising universality
class.  As we now move to a small $1/N_\tau > 0$, 
we may expect the transition point, $\beta_c$, 
to change according to the corresponding finite-size scaling exponents.
More precisely, a finite $aN_\tau$ acts as a length scale cutoff, so
that if the correlation length $\xi$ critical exponent $\nu$ is defined as
$\xi \sim |\beta-\beta_c|^{-\nu}$, we may expect
\be
 |\beta_c(N_\tau)-\beta_c(N_\tau = \infty)| \approx
 \frac{1}{N_\tau^{1/\nu}} \biggr( c_1 + \frac{c_2}{N_\tau^\omega} +
 ...\biggl).  \la{fss}
\ee
Here we have also included $\omega$, the universal 
correction-to-scaling exponent.  
For 3d Ising, $\nu \approx 0.63$, $\omega \approx 0.8$.  The
results, obtained by fixing the exponents $\nu,\omega$ 
and fitting for $\beta_c(N_\tau = \infty), c_1, c_2$, are shown
in~\fig\ref{fig:scaling} for $N_y=2$.  We find that the ansatz in
\eq\nr{fss} indeed allows to fit the data well, as long as
$N_\tau \gsim N_y$.

\begin{figure}[tb]

\centerline{%
    \begin{minipage}[c]{6.4cm}
    \psfig{file=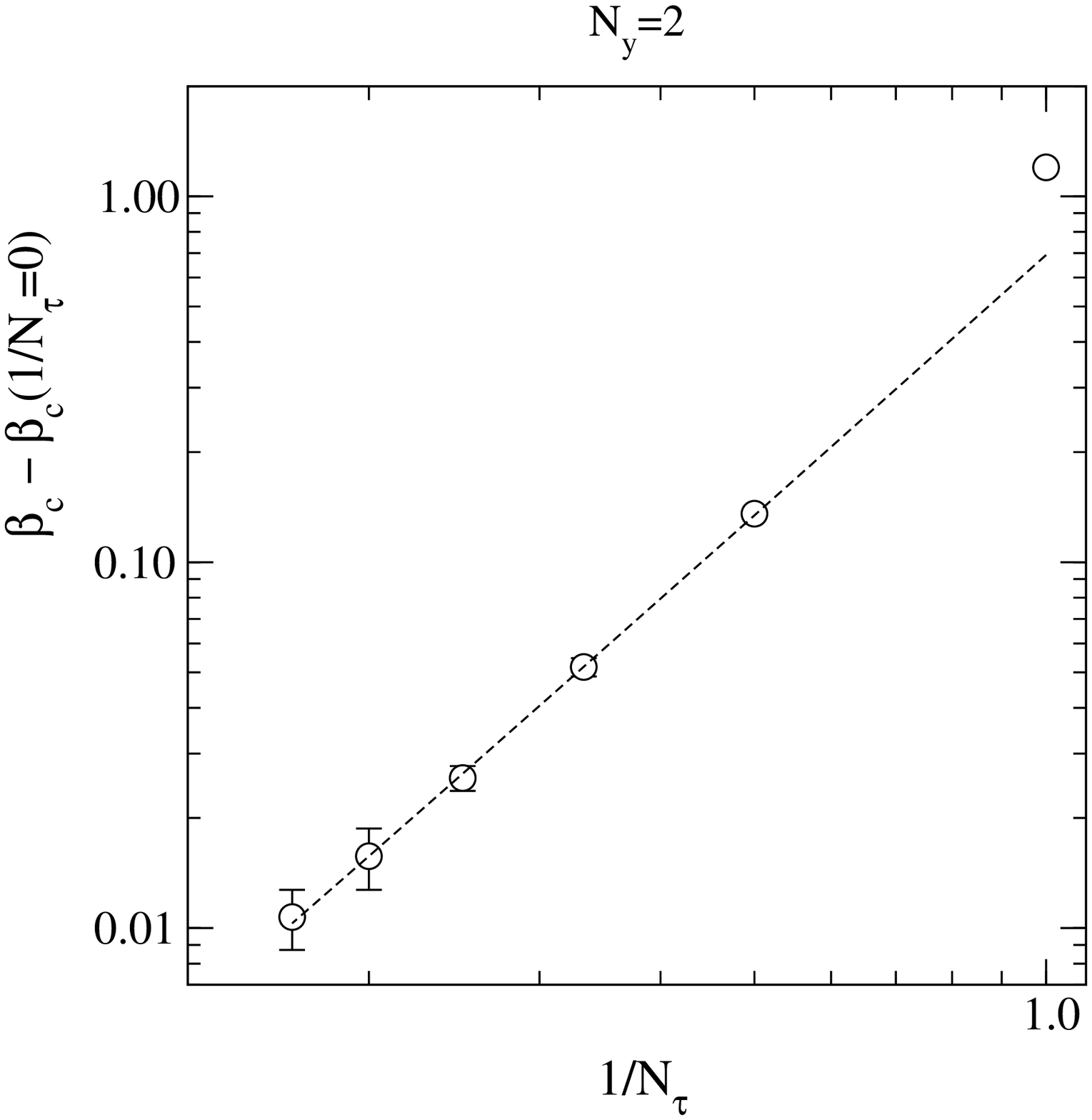,angle=0,width=6.4cm} \end{minipage}}

\vspace*{0.5cm}

\caption[a]{Finite-size scaling of the phase transition 
point determined from $\chi(P_y)$ 
on volumes $16^2\times N_\tau \times 2$,
as a function of $1/N_\tau$,
together with a three-parameter fit 
according to \eq\nr{fss} (fit for $\beta_c(N_\tau = \infty), c_1, c_2$).}

\la{fig:scaling}
\end{figure}

\subsection{Physical implications}
\la{se:physi}

After these consistency
checks, we draw the following physical conclusion
from~\fig\ref{fig:unphys_4d}.  Suppose we fix $\beta_G$, meaning that
we fix the lattice spacing. Staying at this $\beta_G$ at zero
temperature, $1/N_\tau = 0$, let us increase $M$ until we are in the
phase where the symmetry related to the phase of $P_y$ is broken (region C),
by decreasing $N_y$ (recall that $M = 1/(a N_y)$).  This 
effectively guarantees that
dimensional reduction takes place~\cite{adjoint}.  Since $|P_\tau| = 0$,
the dimensionally reduced theory is in its confining phase.  
Increasing now the temperature by moving to the right
($T = 1/(a N_\tau)$), we observe that there is first a deconfinement
transition to region A and then, at large enough values
($1/N_\tau > 1/N_y$), we do cross the transition to the
symmetric phase, $|P_y| = 0$ (region B).
Thus, the phase diagram is indeed as proposed in~\cite{kal}.

\subsection{Phase diagram in continuum units}
\la{phys_4d}

To be clearer with the comparison, we can transform the axes
in~\fig\ref{fig:unphys_4d} to physical units. Through dimensional
transmutation, the 4d SU(2) theory is characterised by a physical
scale which one could choose to be, say, the lightest glueball
mass, the string tension, the Sommer scale $r_0$, or
the scale parameter~$\lambdamsbar$ of the $\msbar$
dimensional regularisation scheme. Here 
it is convenient to set the scale instead through
$T_0$, defined to be 
the critical temperature for the finite temperature phase
transition in the (3+1)d theory with $M=0$.

For the purposes of this paper, it is enough to carry out the
conversion from $\beta_G$, $N_\tau$, $N_y$ to $T/T_0$, $M/T_0$ on a
rather approximate level. The explicit procedure we adopt
is as follows. For $N_\tau = 1,...,16$, the values of $\beta_G$
corresponding to the deconfinement phase transition in the (3+1)d
theory have been determined in~\cite{bems,fhk}. Let us denote these
values by $\beta_c(N_\tau)$. By a spline interpolation\footnote{%
   The spline interpolation introduces a free parameter, the value of the
   second derivative of $\beta_c$ at the boundary $N_\tau = 1.0$, which
   we arbitrarily choose to vanish. On our resolution, this has little
   effect.}, 
we treat the function $\beta_c(N_\tau)$ as if it were
defined for real arguments in the interval $N_\tau = 1.0...16.0$.
We then declare that scaling violations (or finite $a$
corrections) in $T_0$ are small. Consequently, a given value of
$\beta_G$ is converted to a value of $a T_0$ through 
\be
 \frac{1}{a T_0} = N_{\tau,c} \equiv \beta_c^{-1}(\beta_G)\,.
\ee
Given that $a N_\tau = T^{-1}$, $a N_y = M^{-1}$, we then get
\be
 \frac{T}{T_0} = \frac{\beta_c^{-1}(\beta_G)}{N_\tau}, \quad 
 \frac{M}{T_0} = \frac{\beta_c^{-1}(\beta_G)}{N_y}.  
\ee
Whether dependence on the lattice spacing has thus disappeared, 
can be checked explicitly by carrying out the same conversions at
various $N_y$; we do this for $N_y = 2,4$.  On the logarithmic scales
we shall use, the residual dependence turns out to be very small.
To address also values of $\beta_G$ larger than $\beta_c(16.0) =
2.74$~\cite{fhk}, we run $\beta_G$ from there up with the
perturbative 2-loop $\beta$-function.

\begin{figure}[tb]

\centerline{~~\begin{minipage}[c]{9.0cm}
    \psfig{file=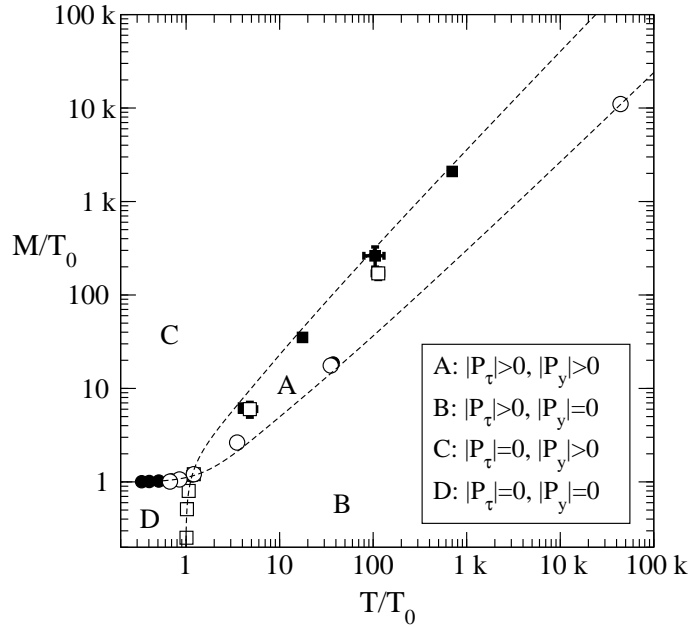,angle=0,width=9.0cm} \end{minipage}}

\vspace*{0.5cm}

\caption[a]{The 4d phase diagram in physical units.
Filled symbols are for $N_y = 2$, open ones for $N_y = 4$. 
That they agree within errorbars indicates 
the absence of significant finite lattice spacing effects. 
The dashed curves represent the finite size scaling 
ansatz shown in~\fig\ref{fig:scaling}, as well as its mirror with respect to 
the interchange $\tau \leftrightarrow y$.}

\la{fig:phys_4d}
\end{figure}

The result of this procedure
is shown in~\fig\ref{fig:phys_4d}. This figure can directly
be compared with \fig2 of~\cite{kal}; we find perfect
qualitative agreement.
It may also be noted how remarkably well the finite size scaling
ansatz in~\eq\nr{fss} can be used to parameterise a fit through the
data points.

In principle it would also be interesting to check the parametric form
proposed for the continuum critical curve (between, say, the regions A,B), 
$T \sim M/g^2 \sim M \ln M$. 
We have however only a few data points, 
with modest resolution and without a precise extrapolation to the 
infinite volume and continuum limits, so we do not attempt that here. 

\section{Lattice results in five dimensions}
\la{se:five}

We then move to 5d. Considerably less can be achieved now, since the
theory is not renormalisable and thus a continuum limit cannot be
systematically approached. Nevertheless, there are a number of issues which
remain the same as in 4d: the 1-loop finite temperature
effective potential for the phase of $P_y$ is still finite 
and shows symmetry breaking at zero temperature 
and small coupling~\cite{bk}, but no sign of symmetry
restoration at high temperatures~\cite{kal}.  At the same time,
non-perturbative arguments can again be presented according to which
restoration should take place. We shall here study the issue in
lattice theories defined with fixed finite lattice spacings.

\begin{table}[t]

\centering
\begin{tabular}{lll}
\hline
 $N_\tau$ & $N_y$ & $N_1\times N_2 \times N_3$ \\ \hline 
 1,2,4,6,8,12   & 1 & $12^3$ \\ 
 1,2,4,8,12     & 2 & $[8^3],12^3,[14^3,16^3,18^3]$ \\ 
 1,2,4,8        & 4 & $8^3,12^3,[14^3]$ \\
\hline
\end{tabular}

\caption[a]{The volumes studied in the 5d case. Note that on account
of the symmetry $N_\tau\leftrightarrow N_y, P_\tau \leftrightarrow
P_y$, some of them are effectively listed twice. The volumes in
brackets have only been used for some of the values of $N_\tau$ shown.}

\la{5d_volumes}
\end{table}

The volumes used are listed in~Table~\ref{5d_volumes}. The action,
observables, and simulation techniques follow closely those in~4d, as
described in~\se\ref{se:four}.

\subsection{Evidence for phase transitions}

\begin{figure}[tb]

\centerline{~~\begin{minipage}[c]{6.4cm}
    \psfig{file=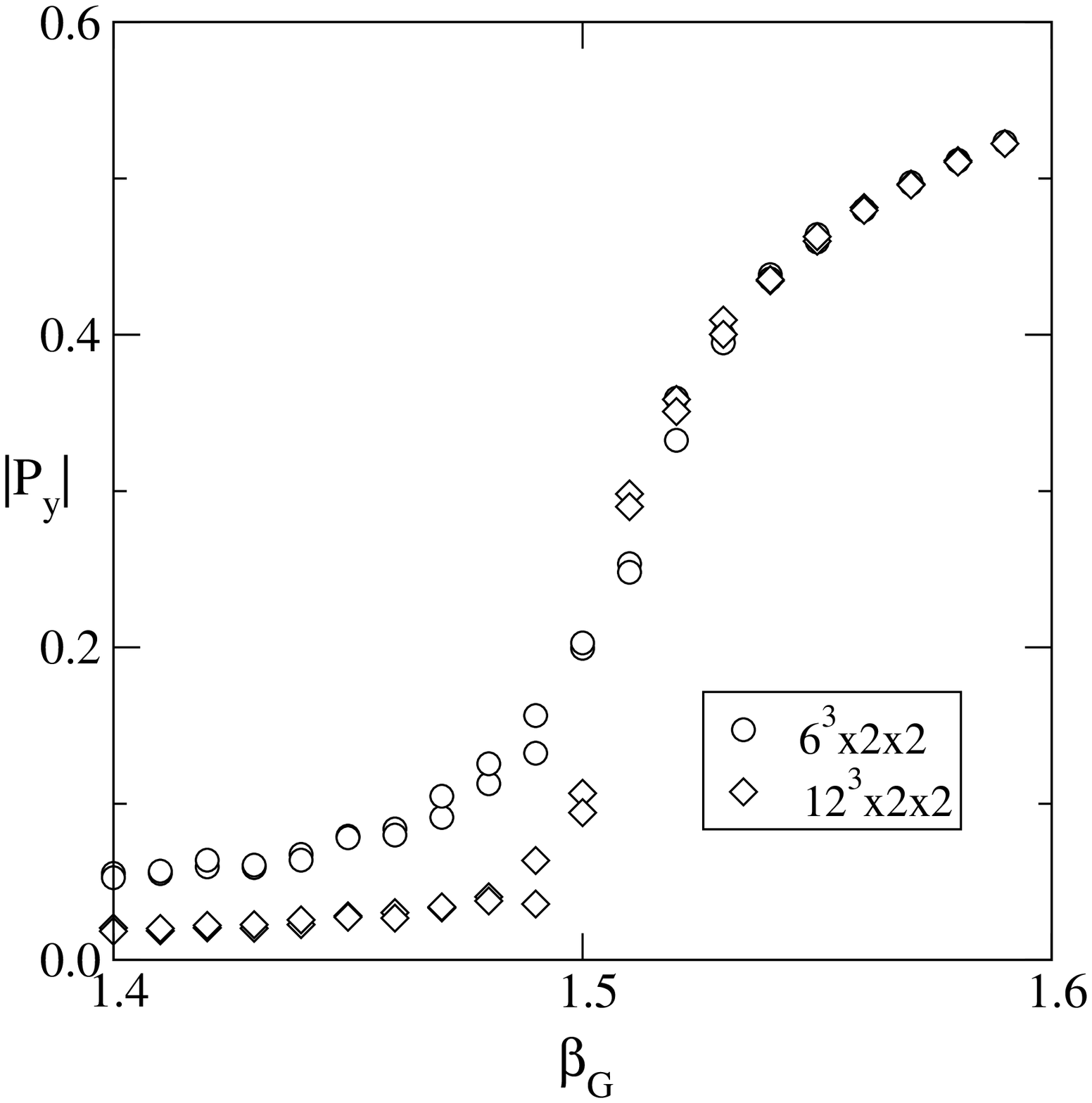,angle=0,width=6.4cm} \end{minipage}%
    \begin{minipage}[c]{6.4cm} \psfig{file=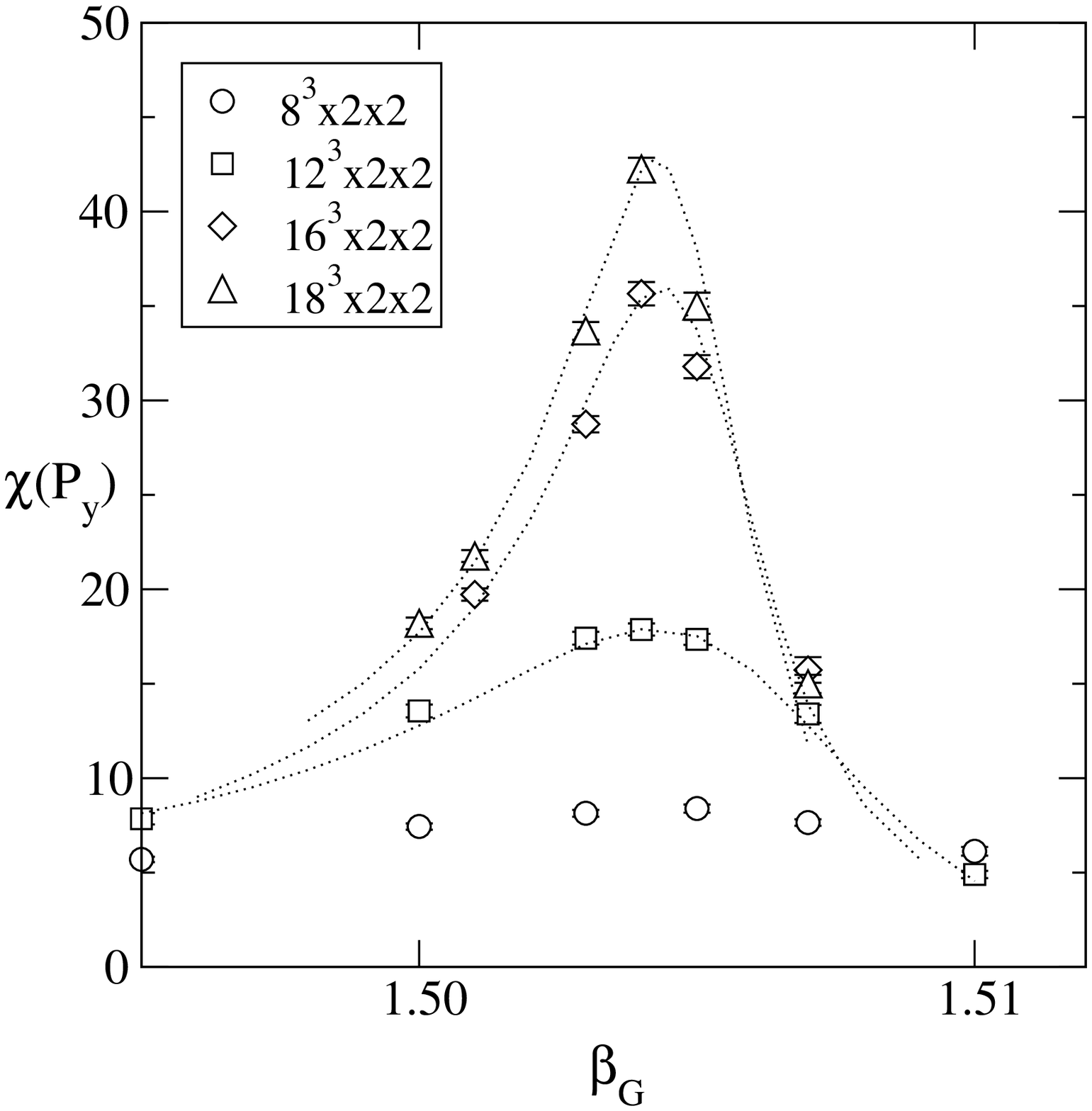,angle=0,width=6.4cm}
    \end{minipage}}

\vspace*{0.5cm}

\caption[a]{Examples of the absolute value 
of the Polyakov loop, $|P_y|$ (left), and the corresponding
susceptibility, $\chi(P_y)$ (right), as a function of $\beta_G$. The
absence of significant
hysteresis on the left indicates a second order phase
transition.  Note that $|P_y|\to 0$ for $\beta_G$ to the left of the 
susceptibility maximum, as the volume is increased. 
The points on the right have been joined together by reweighting.}

\la{5d_second}
\end{figure}

\begin{figure}[tb]

\centerline{~~\begin{minipage}[c]{6.4cm}
    \psfig{file=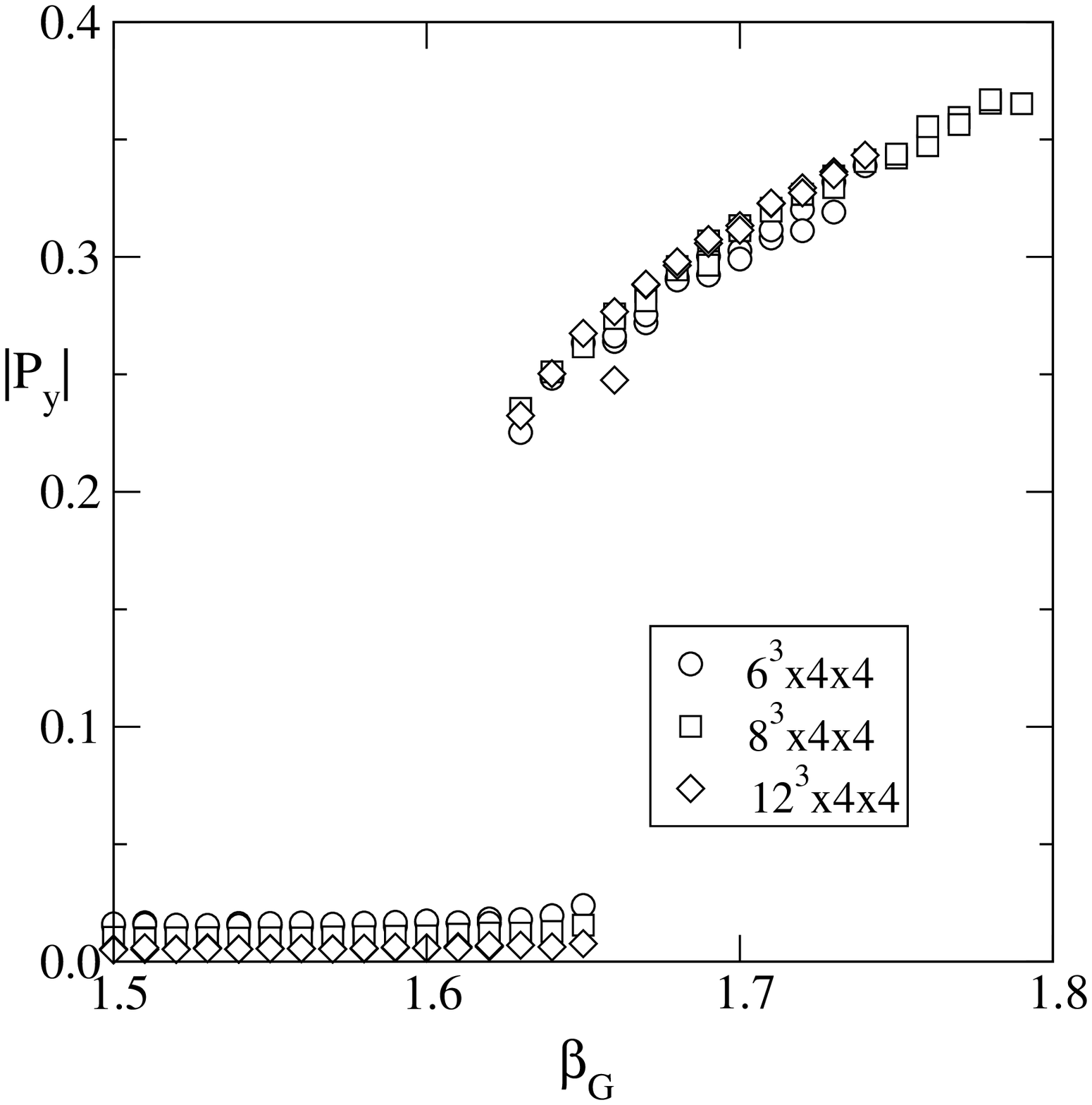,angle=0,width=6.4cm} \end{minipage}%
    \begin{minipage}[c]{6.8cm} \psfig{file=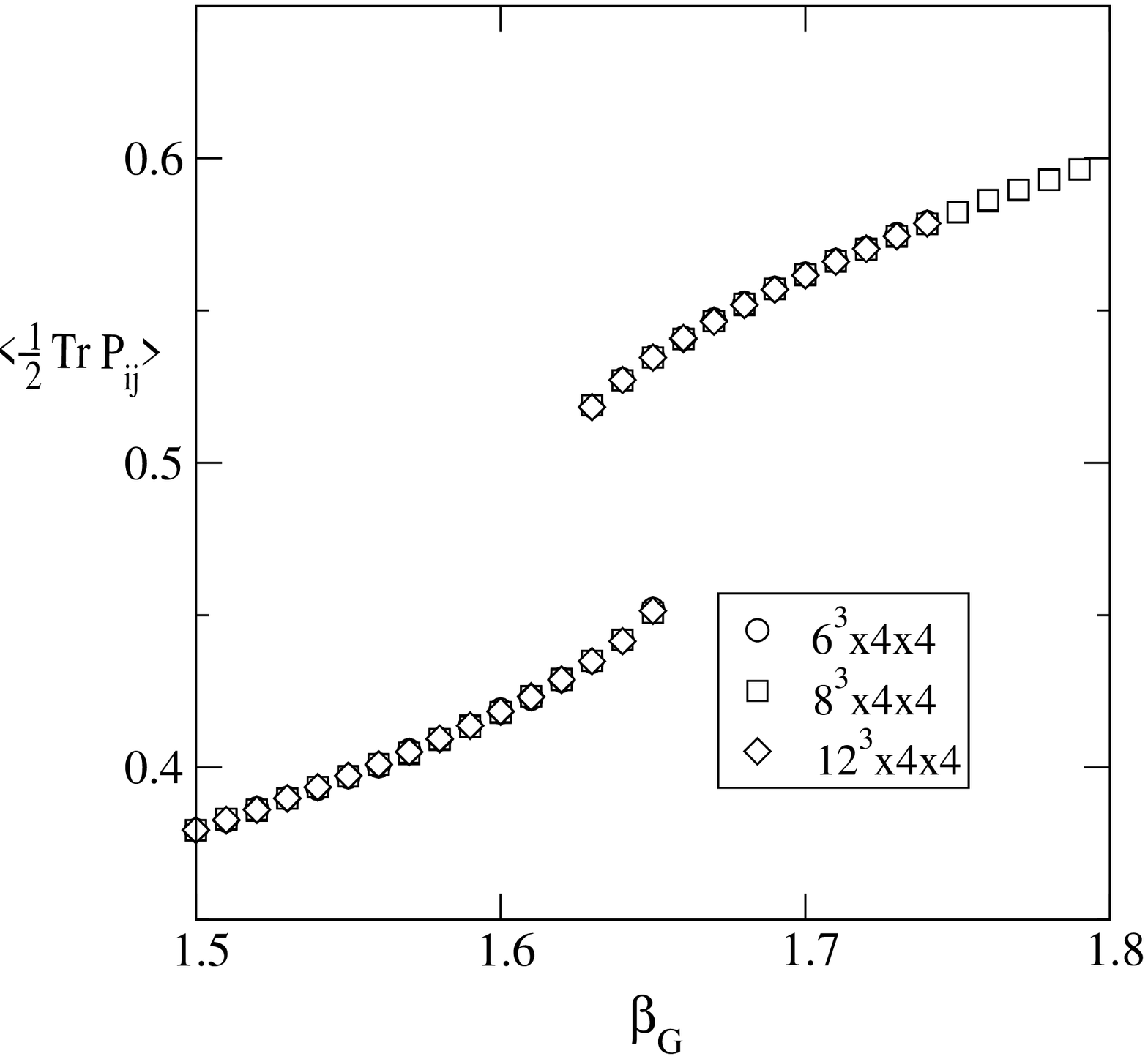,angle=0,width=6.8cm}
    \end{minipage}}

\vspace*{0.5cm}

\caption[a]{Examples of the absolute value of the Polyakov loop, 
$|P_y|$ (left), and the plaquette expectation value, $\langle \half \tr P_{ij}
\rangle$ (right).  The different branches show some metastability, or
hysteresis, indicating a first order phase transition.
Note again that $|P_y|\to 0$ for $\beta_G$ to the left of the 
transition point, as the volume is increased.}

\la{5d_first}
\end{figure}

Typical signals for phase transitions in 5d are shown in
\figs\ref{5d_second}, \ref{5d_first}. In contrast 
to 4d, we now observe both second (\fig\ref{5d_second}) and first
(\fig\ref{5d_first}) order phase transitions.  The first order
transition is also visible in the plaquette expectation value, as
shown in~\fig\ref{5d_first} (right).  Qualitatively, the possibility
of an emergence of a  
first order signal may perhaps be understood via the universality
arguments presented in~\se\ref{se:univ}.

\subsection{Phase diagrams for fixed $N_y$}

\begin{figure}[tb]

\centerline{~~\begin{minipage}[c]{6.4cm}
    \psfig{file=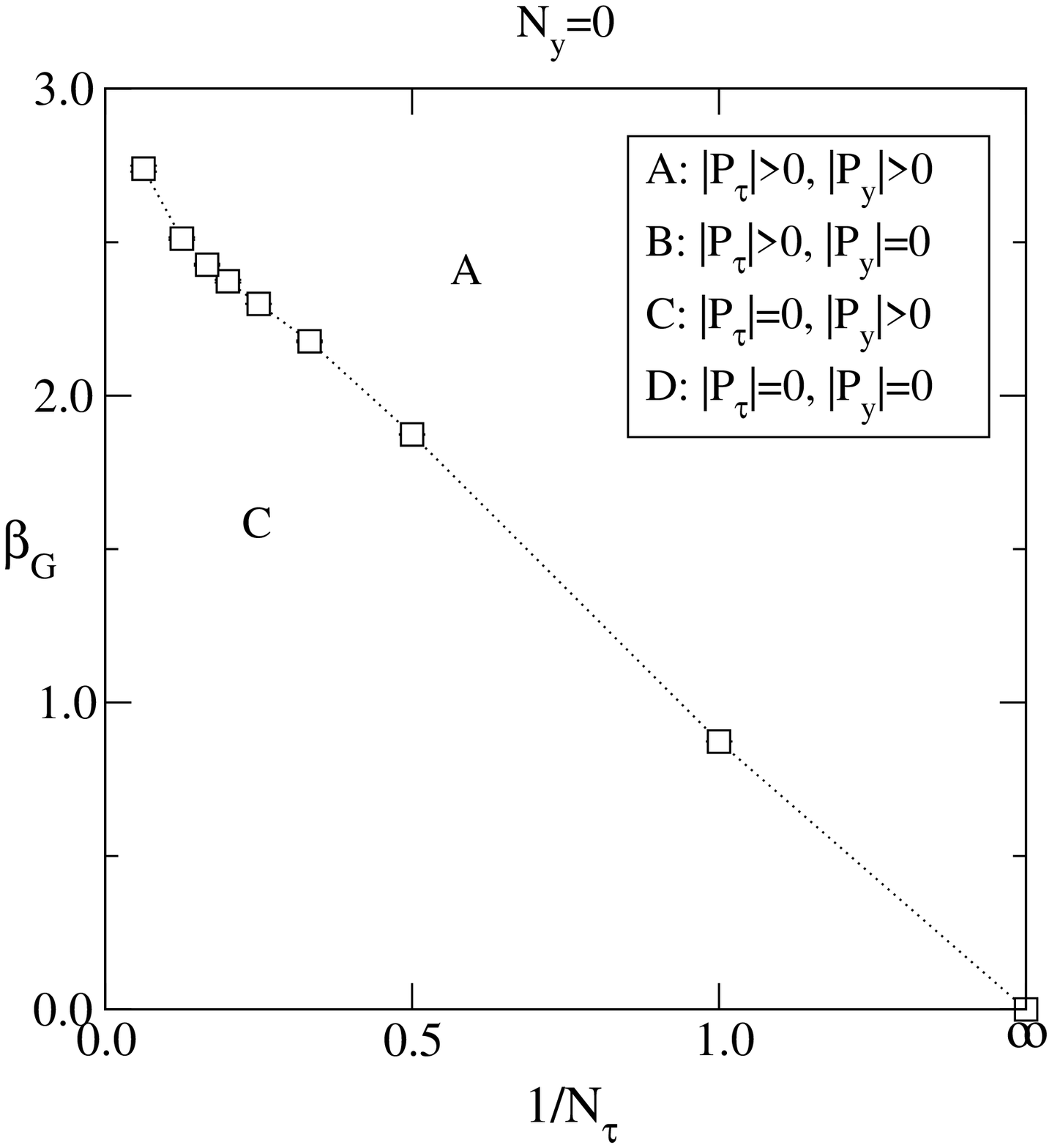,angle=0,width=6.4cm} \end{minipage}%
    ~~~\begin{minipage}[c]{6.4cm}
    \psfig{file=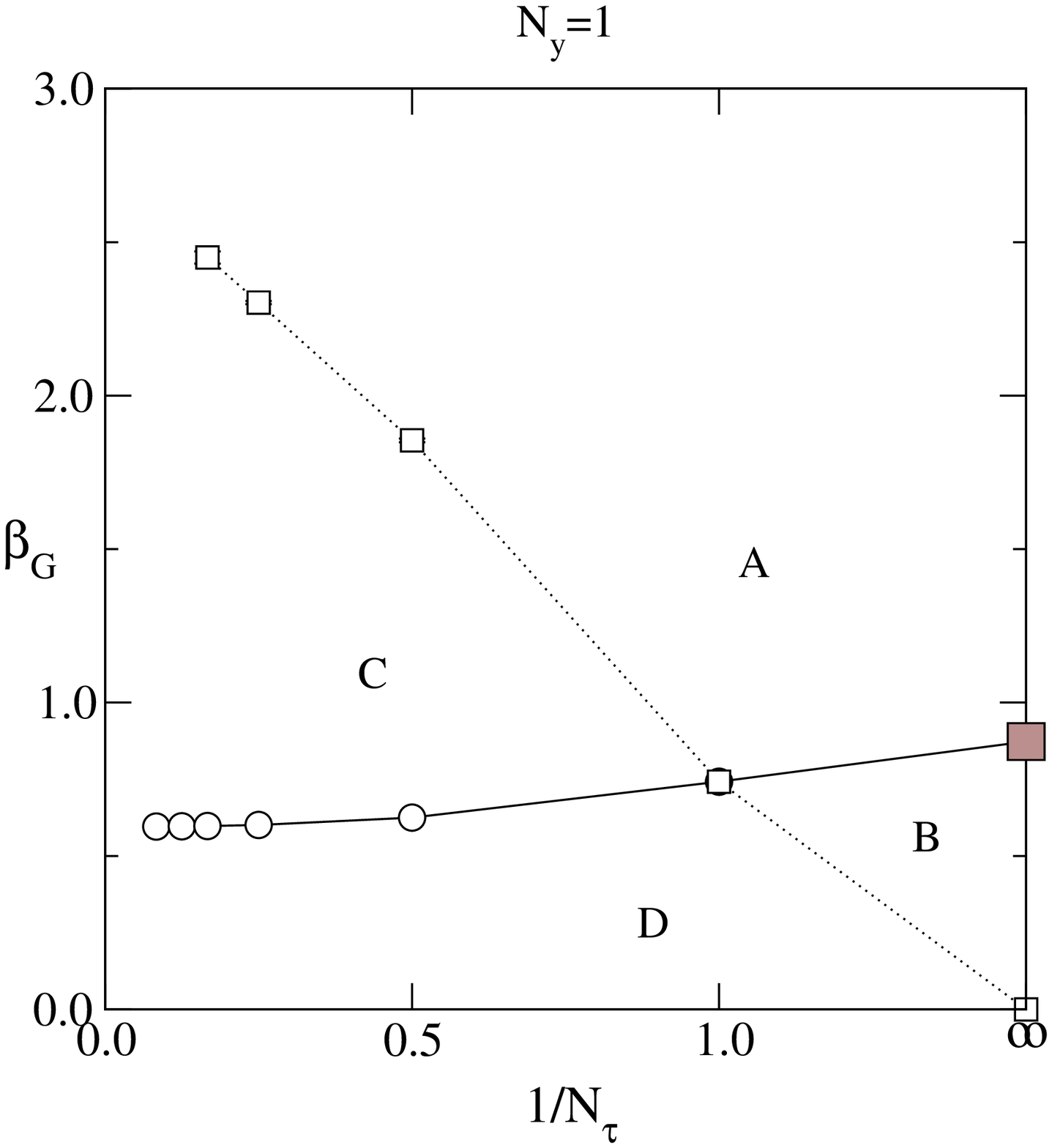,angle=0,width=6.4cm} \end{minipage}}

\vspace*{0.5cm}

\centerline{\begin{minipage}[c]{6.4cm}
    \psfig{file=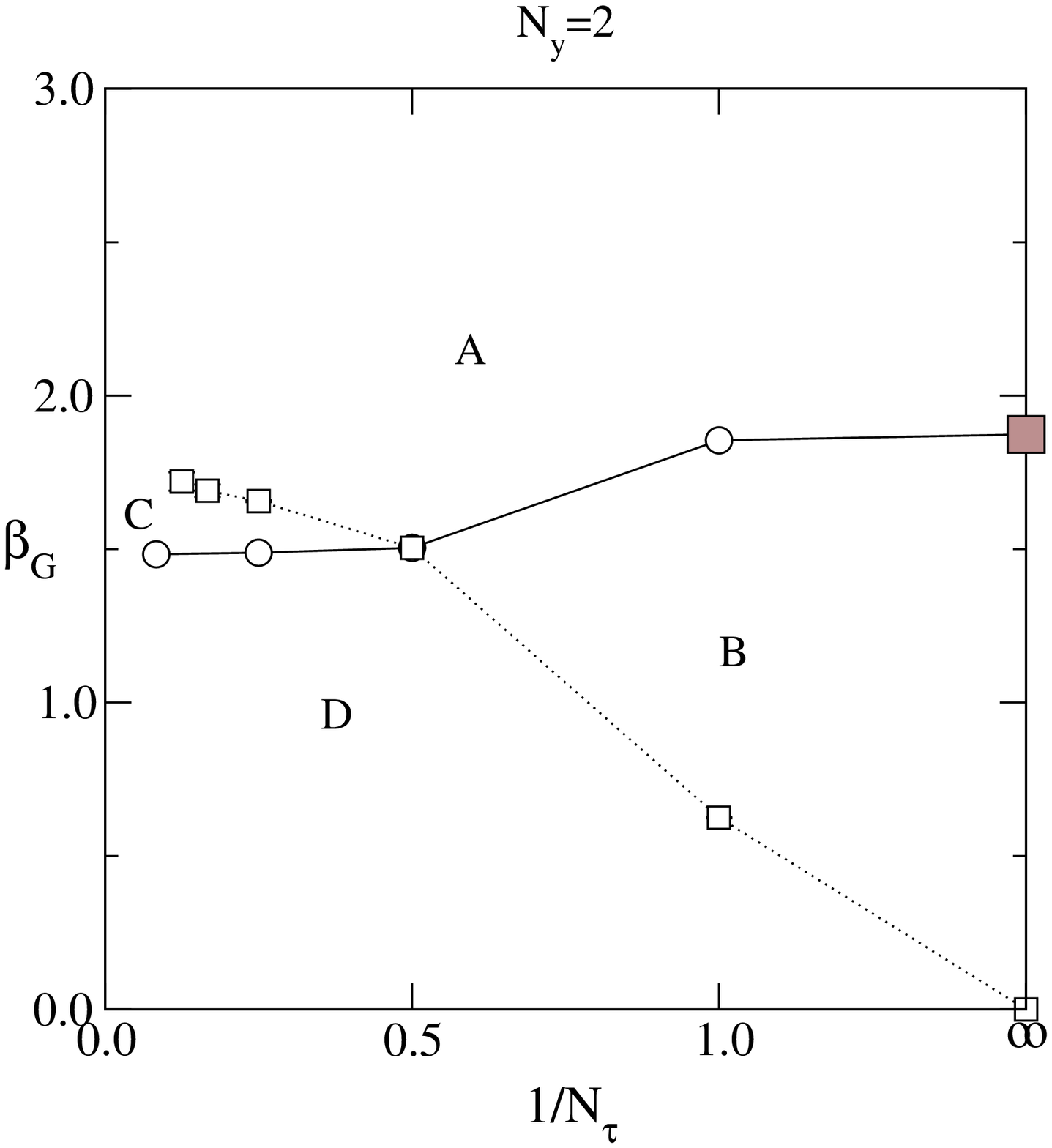,angle=0,width=6.4cm} \end{minipage}%
    ~~~\begin{minipage}[c]{6.4cm}
    \psfig{file=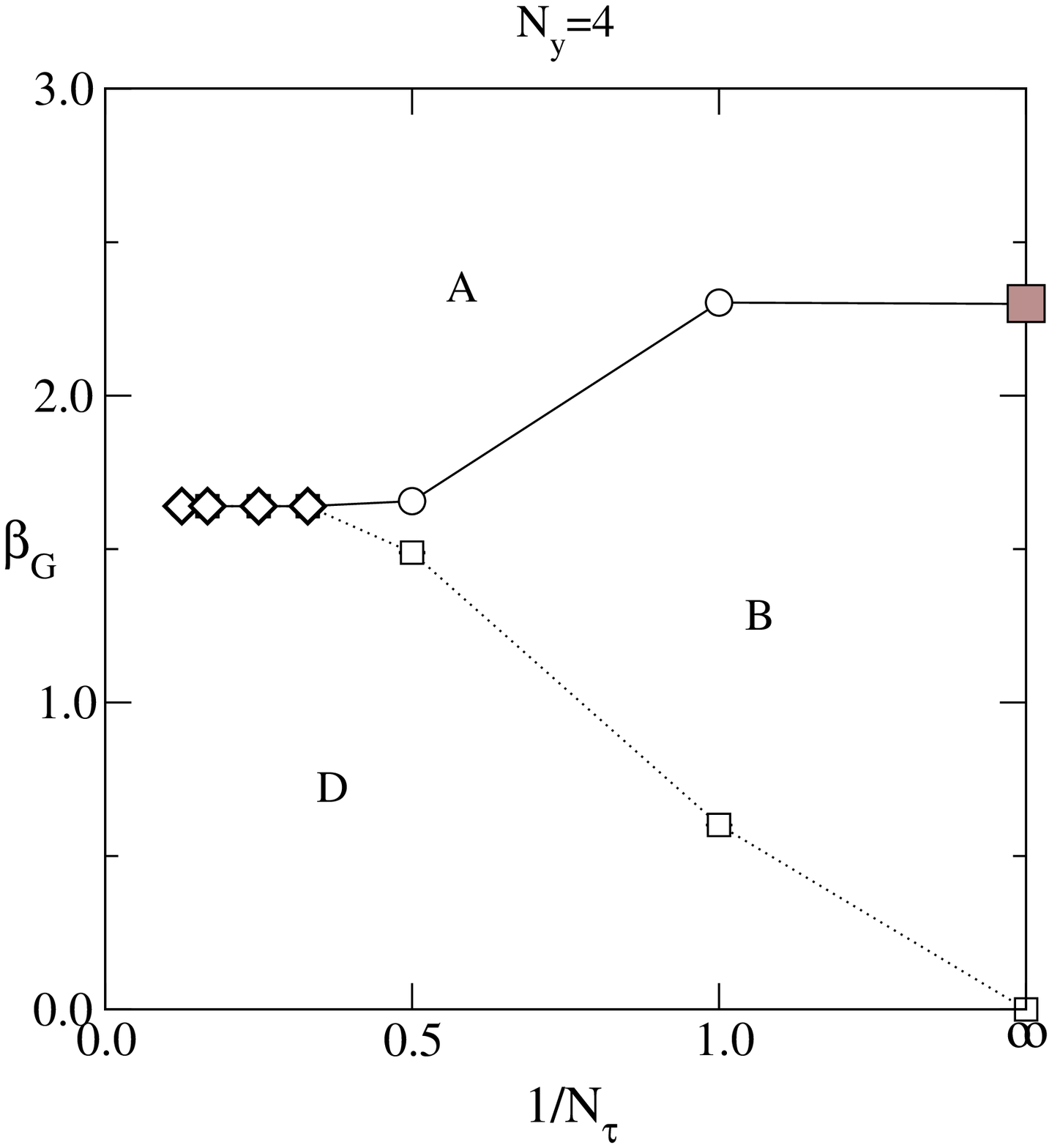,angle=0,width=6.4cm} \end{minipage}}

\vspace*{0.5cm}


\caption[a]{Phase diagrams for various $N_y$, in the 5d case. 
The lines connecting the simulation points are there to guide the eye
only. The values for $N_y = 0$ (top left), as well as for $1/N_\tau =
\infty$ (filled boxes), are
from~\cite{bems,fhk}. Open circles and boxes denote second order
transitions, open diamonds first order ones.}

\la{5d_phasediags}
\end{figure}

\begin{figure}[tb]

\centerline{%
    \begin{minipage}[c]{6.4cm}
    \psfig{file=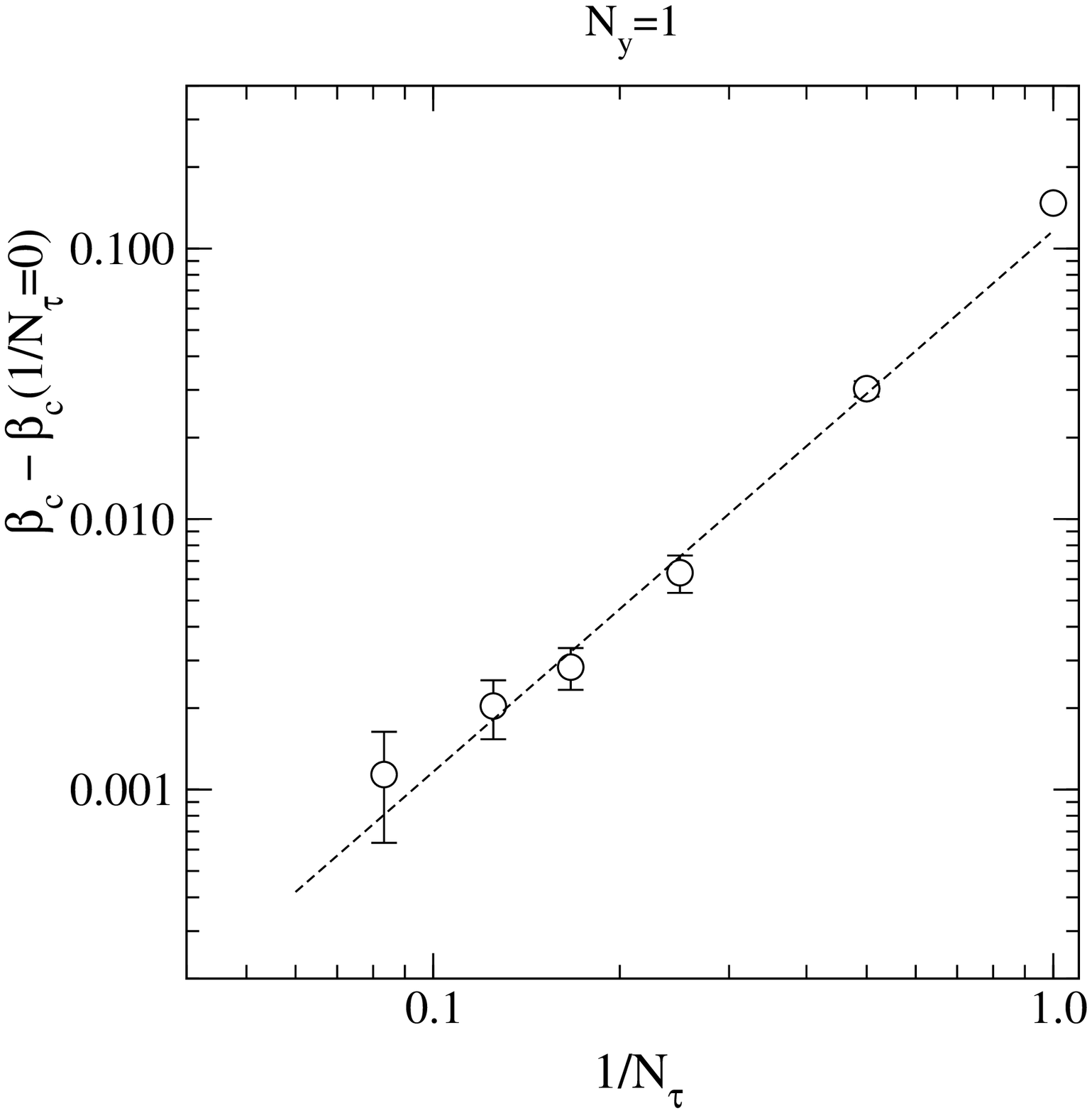,angle=0,width=6.4cm} \end{minipage}}

\vspace*{0.5cm}

\caption[a]{Finite-size scaling of the phase transition 
point determined from $\chi(P_y)$ 
on volumes $12^3\times N_\tau \times 1$,
as a function of $1/N_\tau$,
together with a two-parameter fit according 
to \eq\nr{fss} (fit for $\beta_c(N_\tau = \infty), c_1$,
with a fixed $\nu = 0.5$, $c_2 = 0.0$).}

\la{fig:5dscaling}
\end{figure}

The results from different $N_y$ are summarised
in~\fig\ref{5d_phasediags}.  Again the existing data from the (3+1)d
theory~\cite{bems,fhk} allow for a first check by comparing with a number
of limiting values (filled boxes). We find good consistency.

A second check can be inferred from the scaling of $\beta_c$
close to $1/N_\tau = 0$, as in~\se\ref{unphys_4d}. 
Now the universality class is that of the 4d Ising model, so that
we expect a mean field exponent $\nu = 0.5$ in~\eq\nr{fss}, 
meaning a parabolic
behaviour. In the 5d case our data is not precise enough to address
the issue in great detail, but we do find that, say, the data
with $N_\tau \gsim N_y$ at $N_y = 1$, can be fitted reasonably 
well with two free parameters, $\beta_c(N_\tau = \infty)$ and $c_1$
(\fig\ref{fig:5dscaling}).

\subsection{Physical implications}

Following \se\ref{se:physi}, 
we can then draw the following physical conclusion
from~\fig\ref{5d_phasediags}.  Suppose we fix $\beta_G$, meaning that
we fix the lattice spacing. Staying at this $\beta_G$ at zero
temperature, $1/N_\tau = 0$, let us increase $M$ until we are in the
phase where the symmetry related to the phase of $P_y$ is broken, such
that dimensional reduction to a 4d theory takes place (region C). 
This can be
achieved by decreasing $N_y$, since $M = 1/(a N_y)$. Staying then on
the horizontal line of a fixed $\beta_G$
and increasing the temperature by moving to the right in $1/N_\tau$
($T = 1/(a N_\tau)$), we observe that at large enough values we may
indeed cross the transition to the symmetric phase (region B).  
It is easy
to see that if this happens by passing through
the deconfined region~A, 
then at the transition point $1/N_\tau >
1/N_y$, implying $T_c > M$.  Thus, the phase diagram is
qualitatively as sketched in~\cite{kal}.

However, in contrast to the 4d case, the phenomenon described
only takes place in a limited range of $N_y$, meaning that
the continuum limit cannot be approached. Indeed, we find
(\fig\ref{5d_phasediags}, bottom right) that for $N_y = 4$, the phase
transition close to $1/N_\tau = 0$ is of the first order~\cite{mc},
and the lines corresponding to the breaking of $P_\tau, P_y$ have
merged. This implies that there is no longer a low temperature phase
(region~C)
which would be effectively four dimensional ($|P_y| > 0$) and
confining ($|P_\tau| = 0$)\footnote{ Since we do not address the
Higgs mechanism here, we would like the theory to be in the confinement
phase even if we had in mind gauge fields related to weak
interactions.}.  We expect this to remain true also for $N_y > 4$.
Since the phase diagram changes qualitatively as $N_y$ is increased, 
one cannot define a meaningful line of constant physics, where
discretisation artifacts would gradually vanish, as correlation
lengths measured in lattice units grow.  

We end with a brief comment on a related recent work,
ref.~\cite{eji}.  The authors study a (4+1)-dimensional
SU(2) gauge theory,
corresponding to the axis $1/N_\tau = 0$ in our figures, 
with asymmetric lattice spacings. As far as their
data can be compared with ours, the results are consistent: with our
symmetric lattice spacing, their study corresponds to $N_y \lsim 2$,
implying indeed a second order phase transition. Our conclusion
regarding the existence of a continuum limit
in a genuine 5d theory is however different, for
the reasons mentioned above. Of course a continuum limit does
again exist 
for models of the type in~\cite{moose}, where the extra dimension 
is assumed inherently discretised
(in other words, the continuum limit is only taken on the 4d planes,
will the fifth dimension remains latticized), 
so that the lattice spacing is
in a sense infinitely asymmetric. 

\section{Conclusions}
\la{se:concl}

We have discussed here the finite temperature behaviour 
of pure non-Abelian Yang-Mills theories, living in a space
with a single circular extra dimension.

If the extra dimension is small enough, dimensional reduction takes
place in asymptotically free (or weakly coupled)
theories, as indicated by the standard
perturbative analysis in terms of Kaluza-Klein excitations.  At the
same time, the circular dimension introduces a global Z($N$) symmetry
in the system. If perturbation theory works, the symmetry must 
necessarily be broken, because the expectation value of the Polyakov 
loop in the extra direction is parametrically close to unity. 

While just introduced through perturbative arguments, the same Z($N$)
symmetry breaking can also be observed when the system is 
studied non-perturbatively.
However, as we have demonstrated in this paper,  this symmetry 
gets restored at high temperatures, $T_c > M$, where $M$ is the inverse 
of the perimeter of the extra dimension. This symmetry restoration  
is non-trivial because it is
{\em in contrast} to straightforward perturbation theory. 

Avoiding a domain wall problem in cosmology requires that 
such a symmetry restoration never takes place. We are thus lead to 
conclude that we would either have to change the model, or restrict the
reheating temperature after inflation to be below $T_c$.

The simulations carried out
to reach this conclusion were very simple. In fact, 
for $d = 4$, the limiting cases of zero as well as infinite 
radii of the extra dimension could be extracted from the literature. 
We have here simply added a number of points in between, utilising
small scale numerical simulations. However, 
we believe that our physical conclusions, as discussed above
and summarised in~\fig\ref{fig:phys_4d}, are new. 

It is interesting to note that, at least on the 
resolution of a logarithmic scale, the phase diagram 
in~\fig\ref{fig:phys_4d}
is well described by a finite size scaling ansatz in 
the whole range of interest. It is more 
difficult to determine the parametric behaviour of the 
continuum critical curves, but 
the symmetry restoration temperature is in any case typically above 
the inverse of the perimeter of the extra dimension, $T_c > M$. 

There is less existing data available on the $d=5$ case. The new 
points we have added come again from small scale simulations. On a coarse
lattice, the phase diagram remains qualitatively the same as at $d=4$. 
However, the continuum limit cannot be approached: on a finer lattice, 
some of the phase transitions turn into first order ones, and the 
physically relevant confining phase disappears. 
This means that one cannot define a ``line of constant physics'' 
along which to approach the continuum. Moreover, unlike in 4d, the 
correlation lengths do not diverge in lattice units as $N_y \to \infty$, 
if the first order signal continues to strengthen as expected.
While the absence of a continuum limit is well known, we
may have elaborated on it in a somewhat new way.

For cosmological applications, it is important to discuss the 
robustness of the phase diagram with respect
to small modifications of the model. Let us reiterate that the 
main point of our study is that if a Z($N$) symmetry exists
in the system and is spontaneously broken at zero temperature, 
then it does get restored at high temperatures, contrary to the 
prediction of straightforward perturbation theory. We would 
expect the main model dependence, then, to concern 
the zero temperature starting point, rather than actual finite 
temperature physics. 

It is certainly true that it is possible to break the Z($N$) 
symmetry explicitly with a relatively small modification of the model. 
While matter fields in the adjoint representation do not break it, 
matter fields in the fundamental representation coupling to the covariant
derivative $D_y$, do break the symmetry explicitly. Another generic way 
may be to change the topology of the spacetime, for instance by 
orbifolding, whereby the zero mode of the gauge field 
component $A_y$ vanishes, 
$A_y \equiv -A_y = 0$ and thus $P_y = 1$, 
or by considering more than one extra dimension
of a specific structure. 
At the same time, other topological defects could of course potentially 
be introduced in such more complicated settings 
(see, e.g.,~\cite{lhk,dks}).
To summarise, the existence of domain walls at zero temperature is
by no means generic, but requires a case by case study. 

If the Z($N$) symmetry is indeed explicitly broken by some of the 
mechanisms mentioned, then cosmological constraints are  
much weaker~\cite{vs}. Still, the existence of possibly relatively 
long-lived metastable Z($N$) vacua could lead to interesting phenomena, 
in analogy with the discussion in~\cite{ikr}. 

Let us end by recalling that there are 
other cosmological constraints similar to the
one we have discussed. In particular, the overproduction of gravitons 
(or more generally very weakly coupled Kaluza-Klein excitations)
at temperatures above the inverse radius, in case such particles propagate
in the bulk, requires again a low reheating temperature, $T\lsim M$~\cite{bd}.

\section*{Acknowledgements}

This work was partly supported by the RTN network {\em Supersymmetry 
and the Early Universe}, EU Contract No.\ HPRN-CT-2000-00152,
by the TMR network {\em Finite Temperature Phase Transitions 
in Particle Physics}, EU Contract No.\ FMRX-CT97-0122, 
and by the National Science Foundation, under Grant No.\ PHY99-07949.
We thank D.~Boer for useful discussions. K.F. thanks
the CERN Theory Division for hospitality during a recent visit
where a part of this work was carried out. 


\end{document}